# Tracing Turbulent Ambipolar Diffusion in Molecular Clouds

Hua-bai Li[1], Martin Houde[2], Shih-ping Lai[3] and T. K. Sridharan[4]

[1] *Max-Planck Institute for Astronomy, Königstuhl 17, D-69117 Heidelberg, Germany*
*li@mpia.de*

[2] *Department of Physics and Astronomy, The University of Western Ontario, London, Ontario, Canada N6A 3K7*

[3] *Institute of Astronomy and Department of Physics, National Tsing-Hua University, Hsinchu 30013, Taiwan*

[4] *Harvard-Smithsonian Center for Astrophysics, 60 Garden Street, Cambridge, MA 02138*

**Abstract**

Though flux freezing is a good approximation frequently assumed for molecular clouds, ambipolar diffusion (AD) is inevitable at certain scales. The scale at which AD sets in can be a crucial parameter for turbulence and the star formation process. However, both observation and simulation of AD are very challenging and our knowledge of it is very limited. We recently proposed (Li and Houde 2008) that the difference between ion and neutral velocity spectra is a signature of turbulent AD and can be used to estimate the AD scales and magnetic field strengths. Here we present observational evidence showing that this difference between the velocity dispersions from coexistent ions and neutrals is indeed correlated with magnetic field strength.

## 1. Introduction

Magnetic fields and turbulent flows are ubiquitous in molecular clouds and should be well-coupled at large scales, where ideal magnetohydrodynamics (MHD) is a good approximation. Moving toward small scales, ambipolar diffusion (AD), the decoupling of neutral flows from magnetic fields and plasma becomes inevitable (Mouschovias 1987; Zweibel 2002; Li & Houde 2008). This turbulent AD at small scales might enhance the efficiency of large-scale (mean-field) AD (Zweibel 2002; Heitsch et al. 2004; Fatuzzo & Adams 2002; Nakamura & Li 2005), and the friction between ions and neutrals may accelerate the dissipation of turbulent energy (e.g. Mouschovias & Tassis 2008). Since both turbulence dissipation and mean-field AD can be crucial effects for mass condensation, the turbulent AD scale is an important parameter for the star formation process.

Observing the AD scale directly is difficult, because it may be well below the spatial resolutions of current telescopes. Recently, we (Li & Houde 2008; LH08 hereafter) proposed that turbulent AD can cause differences between the velocity dispersions (VDs) of coexistent ions and neutrals, as measured through spectral line widths, even with telescope beam sizes larger than the AD scale. Moreover, we can use this VD difference to estimate the mean-field strength and the AD scale.

LH08 can readily explain why the velocity dispersions of $HCO^+$ are consistently smaller than those of HCN, as first discussed by Houde et al. (2000 a, b). There are, of course, other mechanisms that may affect the observed VDs or spectral line widths. Especially, optical depths and unresolved hyperfine structures of spectral lines are frequently suspected to play a role in the line width differences between $HCO^+$ and HCN, despite the fact that Houde et al. (2000a, b) and Lai et al. (2003a) had shown that they do not for a significant sample of molecular clouds.

A straightforward way to test the LH08 model is to see whether the difference between the VDs of neutrals and ions is correlated with magnetic field strength, since in their model the difference in VDs is predicted to increase with the strength of the plane of the sky component of the magnetic field. This is exactly what we try to do here. In § 2, we briefly review LH08 and describe our method to test it. § 3 summarizes how the data were obtained. The results are presented in § 4. We discuss other factors that might influence spectral line widths in § 5, followed by a summary in § 6.

## 2. A Test Method of Turbulent AD

### 2.1 *LH08*

An observed turbulent spectral line width is determined by several factors: the intrinsic turbulent power spectrum, the telescope spatial resolution, and the spatial distribution of the tracer within the region probed by the telescope beam. Even when all these factors stay the same for two tracers, difference in optical depths can still make their spectral lines look different.

The main idea of LH08 is that the effect from the difference between coexistent ion and neutral turbulent power spectra at the AD scale is observable even with a telescope beam size significantly larger than the AD scale. This is because a line width results not only from contributions of turbulent eddies with scales comparable to the beam size, but also from all the smaller eddies contained within it. To reveal this subtle effect, one needs a neutral/ion pair to trace very similar volumes of a molecular cloud to minimize influences from other factors.

Houde et al. (2000a, b) showed that $HCO^+$ and HCN have very similar spatial distributions by comparing their maps in a sample of sources. A recent study of G333 cloud by Lo et al. (2009) shows that the degree of correlation between $HCO^+$ and HCN (1-0) maps is the highest (with correlation coefficient 0.94) among many molecular pairs, not to mention neutral/ion pairs.

In Figure 1a we show the $HCO^+$ (4-3) map of M17 (Houde et al. 2002). Some sample lines and the VDs of $HCO^+$ and HCN (4-3) near the peak of the column density (region A) and from the more diffused region (B) are plotted in Figure 1b and 1c respectively. The fact that the VDs from region A are larger is expected for a Komolgorov-type turbulence (e.g. Burkhart et al. 2009, Li et al. 2009), because, while the plane-of-sky (POS) scale is fixed by the beam size, the line-of-sight (LOS) scale of the core region is likely to be larger. More interestingly, the average VD differences between $HCO^+$ and HCN from the two regions are almost identical. LH08 expects the VD differences to be similar as long as the POS magnetic field strengths from the two regions are close (see § 2.2), in spite of the different LOS scales. Varying LOS scales above the AD scale affects VDs in the same way for both ions and neutrals, and thus will not change the VD difference, which stems from AD.

LH08 is supported by recent numerical simulations from both single-fluid treatment (Falceta-Gonçalves et al. 2010) and two-fluid treatment (RIEMANN code; Tilley & Balsara 2010). On the other hand, ion-neutral decoupling does not appear in the two-fluid treatment with "heavy ion approximation" (Oishi & Mac Low 2006). Observational tests are therefore essential.

### 2.2 *The Method*

Also shown in Figure 1a are the POS magnetic field lines (B-lines; see Appendix) based on the polarized dust thermal emission at 350 μm (Houde et al. 2002). The similarity in B-line densities in the two regions is consistent with the assumption that the corresponding POS magnetic field strengths are close to one another. If the ion/neutral VD differences indeed correlate with field strengths as proposed by LH08, then correlation between B-line densities and VD differences should be observed. This will be our test on whether the VD differences stem from turbulent AD. Zeeman measurements are not proper for the test because fields along LOS have no effect on VDs.

The regions we will study are cloud cores and their vicinity, with density $n(H_2)\sim10^{5-6}/cm^3$, which is enough to excite the lower transitions of HCN and $HCO^+$ (Evans 1999).

Submm – mm dust thermal emission are optically thin here (Hildebrand 1983), and their polarimetry detections can not go much lower than this density because of sensitivities of current instruments (Li et al. 2009). So volumes traced by submm-mm polarimetry and by the lower transitions of HCN and HCO$^+$ should be comparable (e.g. Fig.1 of Lai et al. 2003a). HCN and HCO$^+$, however, will be optically thick at and around core peaks, so we need to either use optically thinner isotopologues (i.e. H$^{13}$CO$^+$ and H$^{13}$CN) (§ 3.1), stay away from core peaks (§ 3.2), or estimate velocity dispersions without using the optically thick parts of a line (Fig. 1b).

### 3. Target Selection and Data Acquisition

We need regions with both polarimetry data and HCN/HCO$^+$ detections, as is the case with M17 (Fig. 1). Apart from M17 we also seek clouds with subregions that have similar column densities and, most importantly, significantly different B-line densities.

#### 3.1 DR 21 (OH)

Using the Berkeley-Illinois-Maryland Array, Lai et al. (2003b) presented polarization detections in DR 21(OH) from both the thermal dust emission at 1.3 mm and the CO (2-1) line. These polarization measurements present the POS magnetic field morphology in DR 21(OH). The B-lines based on this work are plotted in Figure 2a. A trend for B-line convergence to the SE is clearly seen.

H$^{13}$CO$^+$ and H$^{13}$CN (1-0) lines from DR 21(OH) were also observed by Lai et al. (2003a) with the Owens Valley Radio Observatory Millimeter Array. They showed that H$^{13}$CN, the wider lines, are optically thin and their hyperfine structures are resolved, so their data are excellent for our test (see § 5). We try to compare the VD differences from regions with low and high densities of B-lines. Figure 3 of Lai et al. (2003a) shows a ridge where the S/N of the VD differences are above 2.5. The northern and southern ends of this ridge, which are correspondingly noted regions A and B in our Figure 2a, happen to be close to the regions with, respectively, the lowest and highest B-line densities.

The position-position-velocity data cubes from Lai et al. (2003a) were processed with MATLAB, where the H$^{13}$CO$^+$ lines and the main hyperfine component of the H$^{13}$CN lines were Gaussian fitted to determine corresponding the VDs using the Curve Fitting Tool.

#### 3.2 NGC 2024

Crutcher et al. (1999) performed Zeeman observations of absorption lines of H I and OH toward NGC 2024 (Figure 2b). The LOS field strength gradually increases from the NE (region A) toward the SW (region B). The peak and minimum of magnetic field strength are symmetrically offset by ~ 1' from the peak of the column density. The C$^{18}$O map from the same work indicates that regions A and B have similar column densities.

Hildebrand et al. (1995) measured the 100 μm polarized emission from the same NGC 2024 area, and the B-lines based on their detections are also shown in Figure 2b. As long as the mean field orientation does not change much (which is very likely; see Li et al. 2009), a correlation between the POS and LOS field strengths should be expected. The B-lines indeed converge on region B.

We have used the SMA (Submillimeter Array; Ho et al. 2004) to measure the VDs of HCO$^+$ and HCN (3-2) from regions A and B, which are centered respectively at (RA, Dec) = (05$^h$41$^m$46$^s$.0, -01°53'12".0) and (05$^h$41$^m$41$^s$.5, -01°54'57".0) (J2000). The observations were conducted on 2008 September 4 (region B) and 8 (region A) with the atmospheric opacity at 225 GHz less than 0.115 (based on the 225 GHz radiometer at the nearby Caltech Submillimeter Observatory). The SMA receivers were operated in a double-sideband mode. Each sideband is 2 GHz wide and comprises of 24 slightly overlapping spectral windows. We set the upper sideband to simultaneously cover the HCO$^+$ and HCN (3-2) lines. Each of the two spectral windows that cover the lines has 512 channels, which provided a velocity resolution of 0.23 km/s. The calibrators for the instrumental gains, spectral bandpass, and absolute flux were, respectively, quasar J0530+135, quasar 3C454.3 and Neptune/Uranus.

The raw data were reduced using the MIRIAD software package. The data cubes were processed with MATLAB, where the lines were Gaussian fitted to determine the corresponding VDs.

## 4. Results

### 4.1 DR 21(OH)

The zero-moment maps (i.e., obtained by the integration of the spectral line profiles) of $H^{13}CO^+$ and $H^{13}CN$ (1-0) are shown in the Figure 1 of Lai et al. (2003a). As they discussed, these maps are highly correlated compared to other species they detected at the same time. The positions at which we compare the VD difference are identified in Figure 2a, and the lines from the strongest and weakest emission from each region are plotted in Figure 3a. The resolved hyperfine components of $H^{13}CN$ are clearly seen.

The VD differences are compared in Figure 3b. The VDs are between 1 and 2.5 km/s. The low-field-strength Region A has a mean VD difference of 0.113 ± 0.073 km/s (one-$\sigma$). The averaged VD difference in the high-field-strength region B is 0.533 ± 0.126 km/s. The variation of the VD difference is 0.42 km/s, more than 4 times the average uncertainty.

### 4.2 NGC 2024

The zero-moment SMA maps of HCN and $HCO^+$ for regions A and B of NGC 2024 are shown in Figure 4a and 4b. It is clear that the maps of the ions and neutrals in Region A (with correlation coefficient 0.77) do not correlate as well as the single-dish data for other sources (e.g. Houde et al. 2000a, b, 2002 and Lo et al. 2009; with correlation coefficient ~ 0.9). The result observed in Region A is not necessarily unexpected, given that the interferometer filters out the larger-scale emission and emphasizes the small-scale intensity fluctuations. The abundances fluctuate differently for the ions and neutrals because they react differently to turbulent shocks (Iglesias & Silk 1978). This effect is especially prominent in the more diffused regions away from cloud cores, like our regions A and B. The situation will not be as serious in a core region, where the core density profile dominates the column density variations of both species, as in the case of DR 21(OH).

Though the abundances fluctuate, the velocity profiles of the two species can still highly correlate as long as they trace the similar volumes of a cloud (see 5.4 for a discussion). Accordingly, where both species are significantly detected in Figure 4, there are always spectral lines (velocity channels) with the correlation coefficient above 0.8 and most of them are above 0.9. We will only compare the VD's of the pairs with the correlation coefficients over 0.9, so that we can be sure that we are looking at neutral and ionized gas that are mostly coexistent in position-velocity space. Note that the velocity profile might be complicated with multiple peaks (Figure 5 a), but there are always channels (11-16 km/s for Region A and 9-15 km/s for Region B) that are highly correlated between the two species.

The positions we have compared are labeled in Figure 4 using lowercase letters. We tried to evenly sample the regions where the zero-moments have S/N > 3 for both species. The spectral lines from these positions are presented in Figure 5a and 5b. To increase the S/N so that the uncertainty on the VDs is reduced, we integrate the emission within a 3″ radius of a position and use the corresponding integrated line profile for the VD.

In Figure 5c we show the the plot of VD(HCN) vs.VD($HCO^+$); the VDs are approximately contained between 0.5 to 1.25 km/s. In the low-field-strength Region A, the averaged VD difference is 0.074 ± 0.029 km/s. Note that the $HCO^+$ is slightly larger in this region (see § 5.1 for discussion). For the high-field-strength Region B, $HCO^+$ is on average 0.133 ± 0.073 km/s smaller than HCN. The variation of the mean VD difference from the low to high field regions is 0.21 km/s, again, more than 4 times of the average of the two corresponding uncertainties.

## 5. Discussion

The VD differences for NGC 2024 and DR 21(OH) indeed appear to correlate with the local field strength. In this section, we discuss other factors that might have influences on line widths, and why we believe that they cannot cause the VD differences we observed. An apparent reason is that none of these factors is expected to correlate with magnetic field strength; more reasons are discussed in the following.

### *5.1 Optical Depth*

Houde et al. (2000b) studied 10 molecular clouds and showed that the $H^{13}CO^+$ and $H^{13}CN$ (J = 4-3 and 3-2) lines are optically thin with the ions slightly more opaque than the neutrals in general. So optical depth broadening cannot systematically explain the smaller VDs of $H^{13}CO^+$. One of their clouds is M17 and they show $H^{13}CO^+$ and $H^{13}CN$ (4-3) detections from the region A and B in Figure 1a.

Assuming that a cloud is homogeneous and isothermal, one can get information about the optical depths of two isotopic variants by comparing their intensities and VDs. The ratio between the optical depths of the thicker and thinner isotopologues is positively correlated with the ratio of their VDs and the inverse ratio of intensities (I) (e.g. Zinchenko et al. 1989). For M17, the ratios $VD(HCN)/VD(H^{13}CN)$ and $I(H^{13}CN)/I(HCN)$ are respectively 1.6 and 0.16 in Region A, and 2.2 and 0.09 in Region B (Figure 1b). Correspondingly, the ratios for $HCO^+$ and $H^{13}CO^+$ are 2.1 and 0.33 for Region A, and 2.0 and 0.27 in region B. So we can expect that for the main isotopologues, the ions are still optically thicker. This analysis agrees with the calculation using the RADEX program (Van der Tak et al. 2007), which assumes non-LTE molecular radiative transfer in an isothermal homogeneous medium. For this we assumed $n(H_2)=10^5$ cm$^{-3}$ and $T_K=20$ K for NGC 2024 (Crutcher et al. 1999), and a typical line width of 3 km/s. The column density of the molecule is uncertain but should be between $10^{12}$-$10^{14}$ cm$^{-2}$ with HCN a factor of two higher (Lucas & Liszt 1996). The optical depths of $HCO^+$ and HCN (3-2) resulting from this calculation are plotted in Figure 6.

Lai et al. (2003a) showed that $H^{13}CN$ (1-0) has three hyperfine lines with intensity ratios of 1 : 5 : 3 for the optically thin case. This also disfavors optical depth broadening as the cause of the wider neutral line widths in DR 21(OH).

Both the optical depths and the fact that $H^{12/13}CO^+$ has slightly larger spatial extent (and thus larger LOS scale) than $H^{12/13}CN$ (Houde et al. 2000a, b; Lai et al. 2003a) should cause the line widths of $H^{12/13}CO^+$ to be wider when the magnetic field is very weak. And this is exactly what we observed in the Region A of NGC 2024 (Figure 5c). This also makes a magnetic field strength estimated by the LH08 method only a lower limit if $HCO^+$ and HCN are used.

### *5.2 Spectral Hyperfine Structure*

If the VD differences between $HCO^+$ and HCN were caused by the different and unresolved separations of their hyperfine components, the VD differences should not vary from regions to regions, contrary to what we see in NGC 2024 and DR 21(OH).

The DR 21(OH) data offer one more evidence against hyperfine structures, because they are resolved for the $H^{13}CN$ lines but not for the narrower $H^{13}CO^+$. Though not the purpose of this work, when estimating field strength, one should keep the unresolved $H^{13}CO^+$ components in mind (which is 0.13 km/s; Schmid-Burgk et al. 2004).

### *5.3 Outflow*

The line width of a tracer sensitive to outflows may be wider because of the enhancement of the red and/or blue wings. It indeed has been reported that $HCO^+$ and HCN emission is enhanced near outflows (e.g. Hogerheijde 2002; Rawlings et al. 2004), but there is no evidence showing any one between the two to have systematically higher enhancement.

The regions A and B near FIR 2 & 3 of NGC 2024 are far away from the outflows present in the cloud, which are likely powered by FIR 5 or 6 (Buckle et al. 2010). Moreover, even if there was evidence showing outflows in the M17 regions we observed, it is hard to argue how the outflows can affect regions A and B, one within the cloud core and one ~2' away, in the same way to produce the similar VD differences (Figure 1c).

### 5.4 Map Morphology Difference

$HCO^+$ and HCN is the ion/neutral pair with the highest spatial correlation (see 2.1). Even so, some differences between their maps still exist, especially at smaller scales, as can be observed in Figure 4.

There are two possible reasons for the map morphology discrepancy. First, the tracers may simply trace different regions of a cloud; second, the same region is traced but the abundances vary in different ways between the tracers from positions to positions. For the latter case, though the intensity ratio varies, the line shapes, e.g. central velocities and VDs, still correlate and thus VD comparisons will be valid.

The highly correlated channels shown in Figure 5 (11-16 km/s for Region A and 9-15 km/s for Region B) suggest that the second scenario dominates for these channels. The intensity fluctuations, however, introduce fluctuations to the optical depths and LOS scales (due to the limited telescope sensitivity), and thus fluctuations in the VDs. But these *fluctuations* can only affect the dispersions about the mean VD, not *systematic* differences between the VDs. Note that the map correlation coefficients are, respectively, 0.77 and 0.89 for NGC 2024 A and B, while the corresponding VD differences are 0.07 and 0.21 km/s. So the morphology differences cannot be the reason of the VD differences we observed.

### 6. Summary

Using $HCO^+$ and HCN lines, we try to trace turbulent ambipolar diffusion, which is supposed to differentiate the velocity dispersions between coexistent ions and neutrals. Furthermore, this difference should be positively correlated to the POS magnetic field strength (Li and Houde 2008). M17, NGC 2024 and DR 21(OH) have been studied; the POS magnetic field line density is pretty uniform for the first and varies significantly for the other two. The results are encouraging: the velocity dispersion difference between ions and neutrals stays close to constant in M17, while the regions with stronger magnetic fields in NGC 2024 and DR 21(OH) have the larger differences in velocity dispersion. We considered other possible interpretations that could result from differences of various parameters between the two tracers, including optical depth, outflow enhancement, and hyperfine structures, but only turbulent ambipolar diffusion can explain our observations.

H.L. appreciates the encouragement and valuable comments from Roger H. Hildebrand, Philip C. Myers, Eric R. Keto and Qizhou Zhang on LH08. We appreciate the help from the SMA staff on observations and from Alice Argon on data analysis. H.L.'s research is supported by the postdoctoral fellowships from Max-Planck-Institut für Astronomie and from Harvard-Smithsonian Center for Astrophysics. M.H.'s research is funded through the NSERC Discovery Grant, Canada Research Chair, Canada Fund programs for Innovation, Ontario Innovation Trust, and Western's Academic Development Fund programs. The Caltech Submillimeter Observatory is funded through the NSF grant AST-0540882. The Submillimeter Array is a joint project between the Smithsonian Astrophysical Observatory and the Academia Sinica Institute of Astronomy and Astrophysics and is funded by the Smithsonian Institution and the Academia Sinica.

# Appendix

Polarimetry maps are usually displayed with discrete vectors with spaces equal to the pixel size of the Stokes Q and U maps (e.g., Figure 2b). When using polarimetry data to infer the orientation of magnetic fields, it is hard to visualize the variation of the field line (B-line) density. Here we develop a method to draw B-lines from the Stokes Q and U maps. Starting with a pixel in the region of interest, a polarization vector can be drawn from the center using the Stokes Q and U values of this pixel, as exemplified with the double-headed vector shown in Figure 7. We use the Q and U values only to determine the orientation, and fix the vector length to be the same as the pixel size. A box with the same size as a pixel centered at one end of the vector (the dashed box in Figure 7) overlaps with the nearby four pixels. The Q and U values at the center of this box is defined as an average of these four pixel values weighted by their overlapped areas with the box. Then a new polarization vector can be drawn from the center of the box with length defined as half the pixel size. Repeating the process, vectors can be drawn head to tail and form a B-line eventually reaching the boundary of the Q/U maps. Starting with the other end of the first double-head vector, we can extend the B-line to the opposite direction.

The grey line plotted in Figure 7, called the "starting line", is perpendicular to the first double-headed vector. On this line, we picked a proper distance D to start the two nearby B-lines, which are defined by the means of nearby pixels in the same way as the first B-line. From the starting point of a new B-line, the starting line is extended in a direction perpendicular to this new B-line by the same distance D, where next B-line will start. Repeating this process, we get a group of B-lines extending out from the region of interest, and the B-line density can be defined by the number of the B-lines and the length of the starting line.

The B-line density in other regions of the map can be calculated by drawing perpendicular lines to the same set of B-lines. One example is shown if Figure 2b, where the longer white dashed line is the starting line, and the shorter white dashed line in the lower-right corner is drawn for the B-line density of Region B.

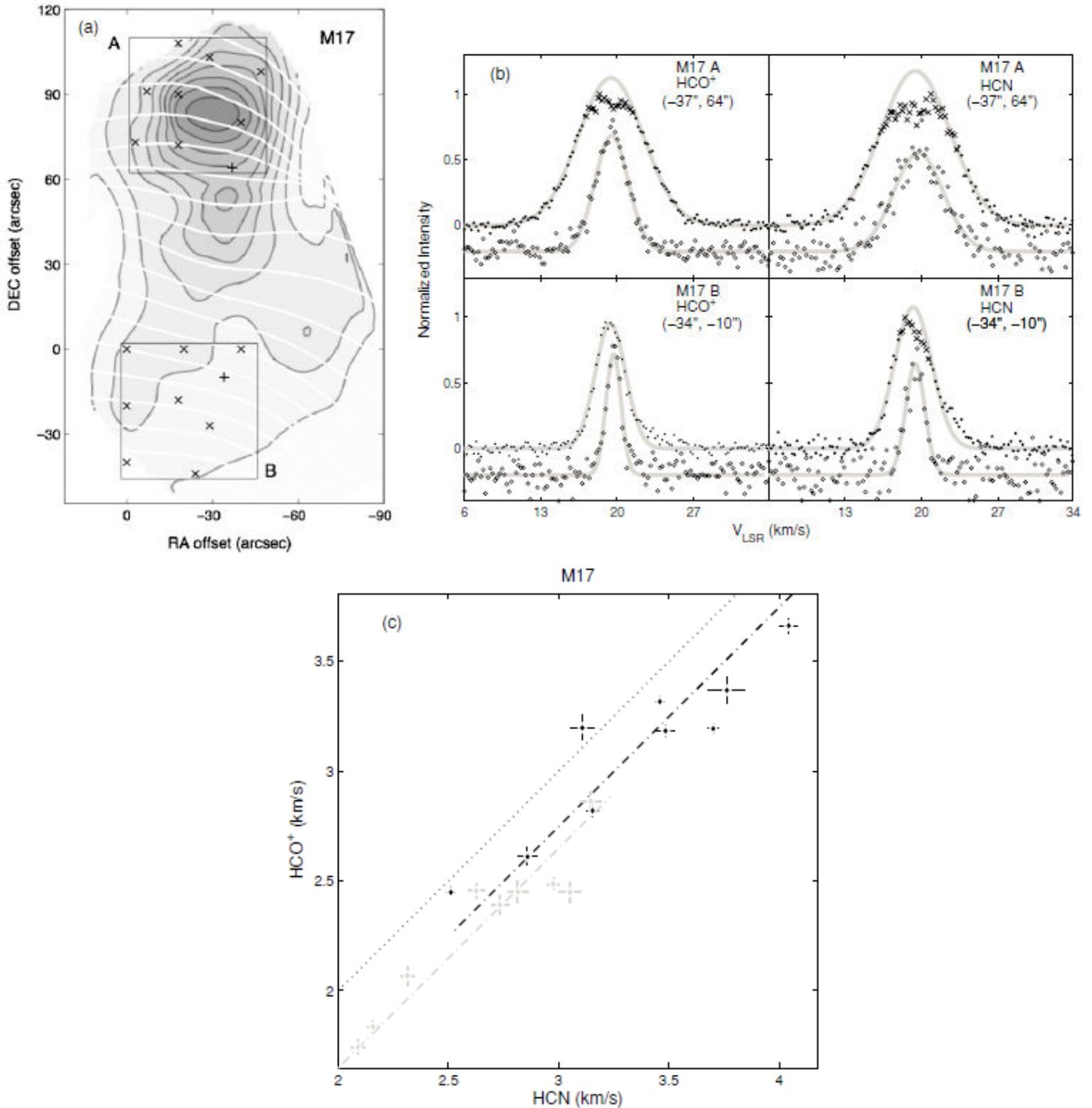

**Figure 1** (a) HCO$^+$ (4-3) map of M17 (Houde et al. 2002). The lowest contour level has Ta*=24 K km/s and the following levels increase linearly with 12 K km/s. The origin is at (RA, Dec) = (18$^h$17$^m$31$^s$.4, -16°14′25″.0) (B1950). The beam size is 20″. The white lines are the magnetic field lines (B-lines; see Appendix) based on polarimetry data of 350 μm dust emission. The square regions A and B have similar B-lines densities but very different column densities. The positions marked by "x" or "+" inside the squares are where we compare velocity dispersions (see text). The positions with "+" have two isotopic variants detected (i.e. H$^{12/13}$CO$^+$ and H$^{12/13}$CN).
(b) The normalized line spectra and their Gaussian fits of H$^{13/12}$CO$^+$ and H$^{13/12}$CN from the "+" positions in Fig 1a. The optically thinner isotopologues are shown as "o"s, and the optically thicker isotopologues are shown by ". "s and "×"s. The "×"s are excluded for Gaussian fits because of their high optical depths. The VD ratios and intensity ratios between these isotopologues give information about their relative optical depths (see section 5.1).
(c) HCO$^+$ vs. HCN on velocity dispersion. The values are from the Gaussian fits to the spectral lines, with the 90% confidence bounds shown by the bars. The dark and light colors are for, respectively, regions A and B. The dotted and dash-dot lines stand for, respectively, the "y = x" line and "y = x-b" fitted to each region. Note that the fitted lines are very close to each other (within 1-σ), regardless of the very different column densities from A and B.

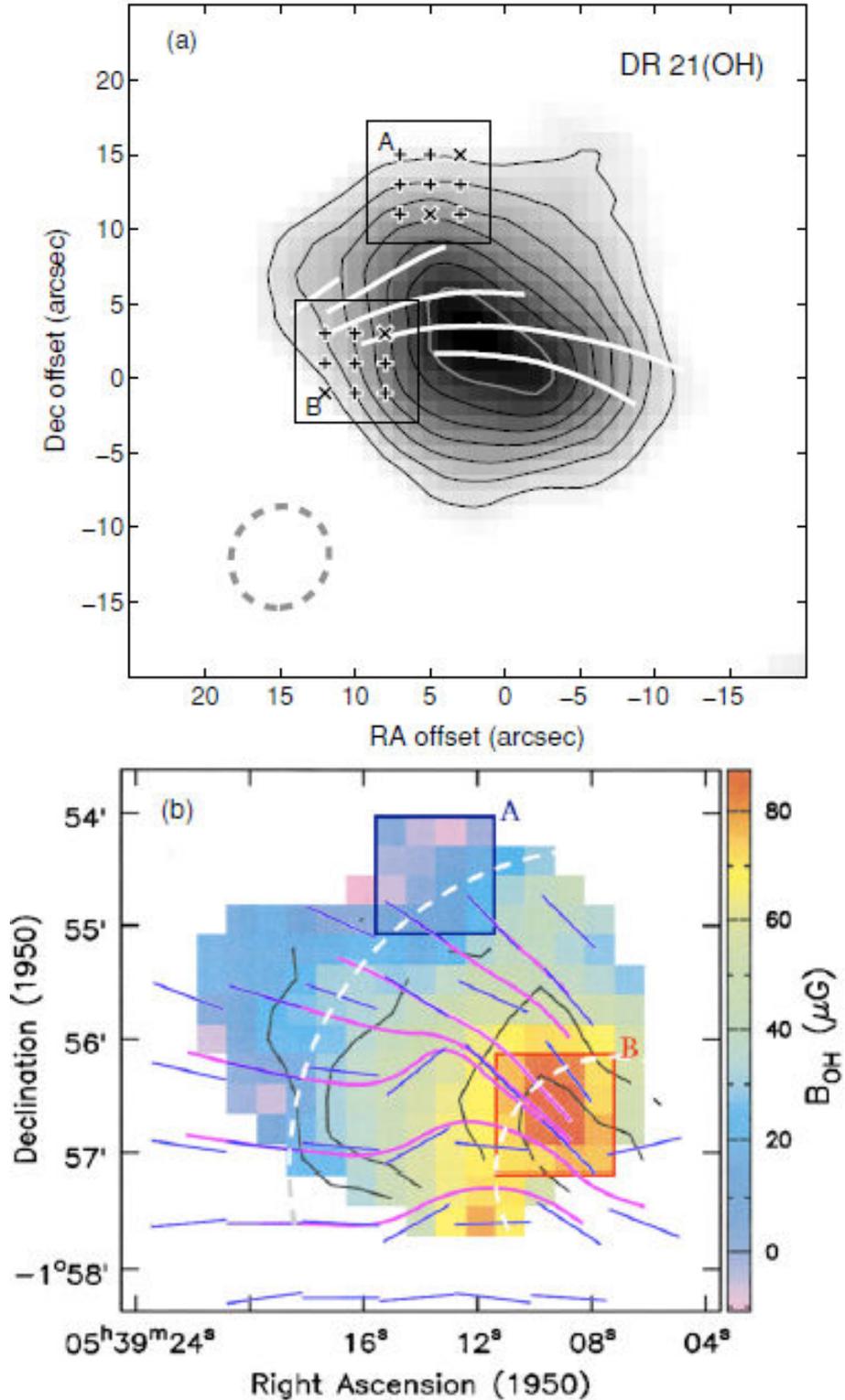

**Figure 2.** (a) H$^{13}$CO$^+$ (1-0) map of DR 21(OH) (Lai et al. 2003a). The origin is at (RA, Dec) = (20$^h$39$^m$00$^s$.72, 42°22′46″.7) (J2000). The peak is 5.2 K km/s and the contours are 90, 80, …, 30 % of the peak value, all with S/N above 3. The dashed oval in the lower-left corner indicate the beam size 6.9″ x 6.4″. The white lines are the magnetic field lines (B-lines; see Appendix) based on the 3mm dust and CO emission. The squared regions are chosen to study the effect of magnetic fields on velocity dispersion. The squares are located in regions with very different B-lines densities, but similar column densities. The positions marked by "x" or "+" inside the squares are where we compare velocity dispersions.

(b) Maps of the line-of-sight magnetic field strength in NGC 2024, estimated by OH (color image) and H I (contours) Zeeman measurements. Contour levels are 15, 30, and 45 µG (adopted from Crutcher et al. 1999). The blue vectors indicate the magnetic field directions inferred from the 100-µm polarimetry data (Hildebrand et al. 1995). The magenta lines are the magnetic field lines (B-lines; see Appendix) based on polarimetry data.

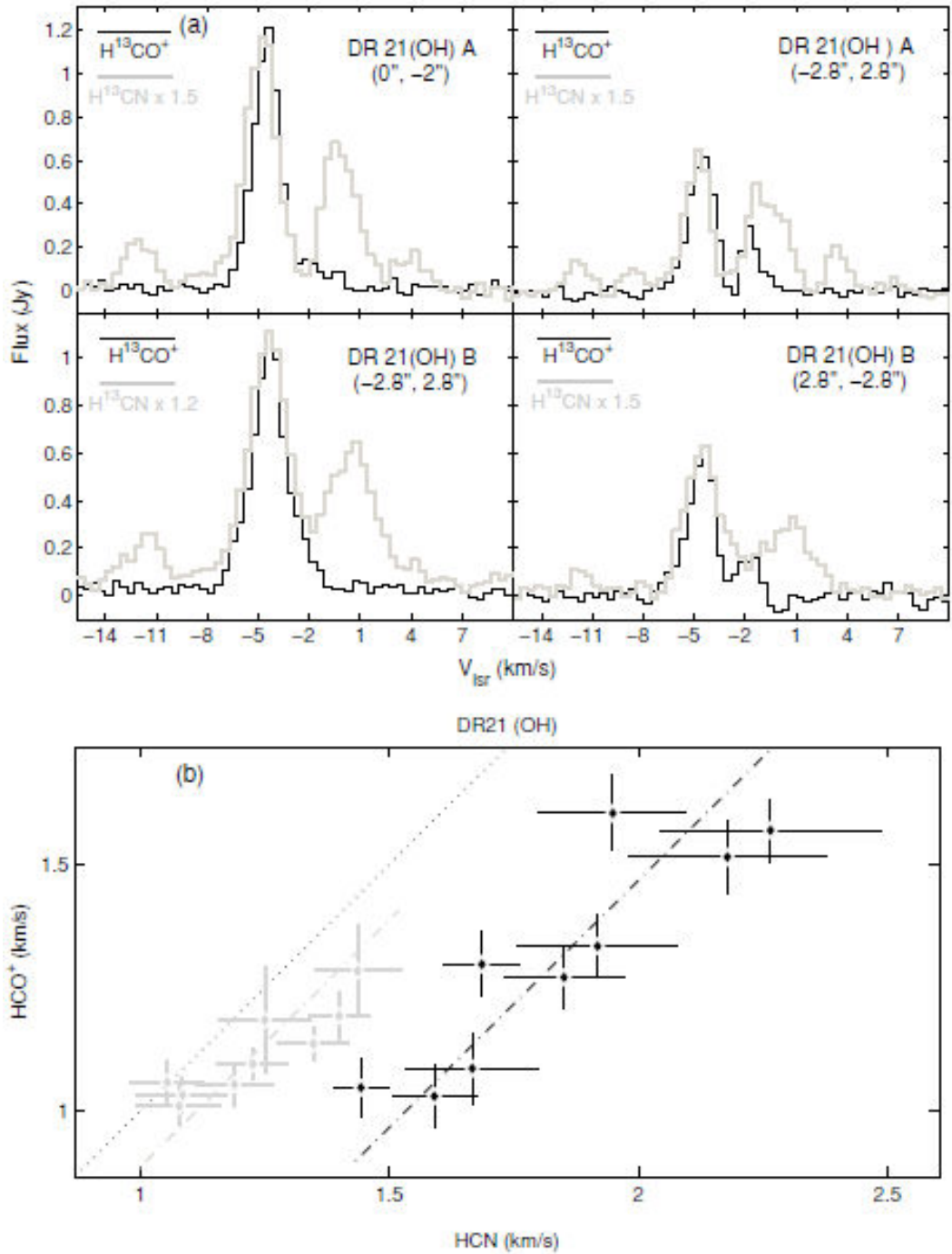

**Figure 3.** (a) Spectral lines of the strongest and weakest emission in DR 21(OH) from the squared regions in Fig. 2a (marked by "×"s). Note that cases have the main (brightest) components well aligned between the ion and neutral, a general condition of all the lines we study for VDs.
(b) Similar to Fig. 1c, HCO$^+$ vs. HCN on velocity dispersion for DR 21(OH). Note how significantly separated the two fitted lines are compared to Fig. 1c.

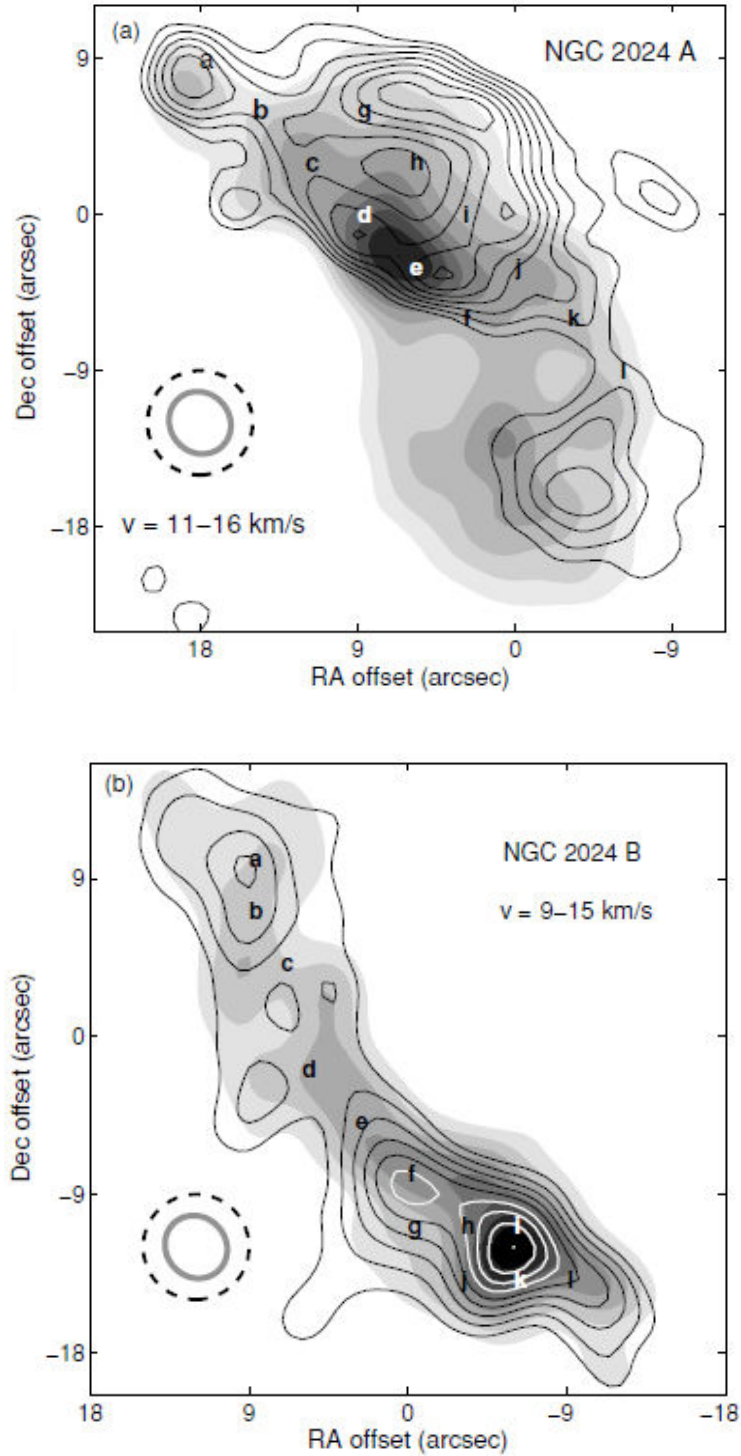

**Figure 4.** (a) & (b) Zero-moment maps of HCO$^+$ (3-2) (filled contours) and HCN (3-2) (dark contour) from, respectively, Regions A and B in NGC 2024. The highest contour level has 90% of the peak intensity, and the following levels decrease linearly by 10 % of the peak value, all with S/N above 3. The highest contour levels are 1.4 (dark) and 4.5 (filled) Jy/(beam km/s) for region A and 8.3 (dark) and 21.0 (filled) Jy/(beam km/s) for region B. The beam sizes are about 3.7″ × 3.4″, as indicated by the ovals in the lower-left corners. The dashed circles with 3″ radius show the size of the areas we integrate to get spectral lines for determining the velocity dispersions (VD), and the areas are centered at positions marked by the lowercase letters.

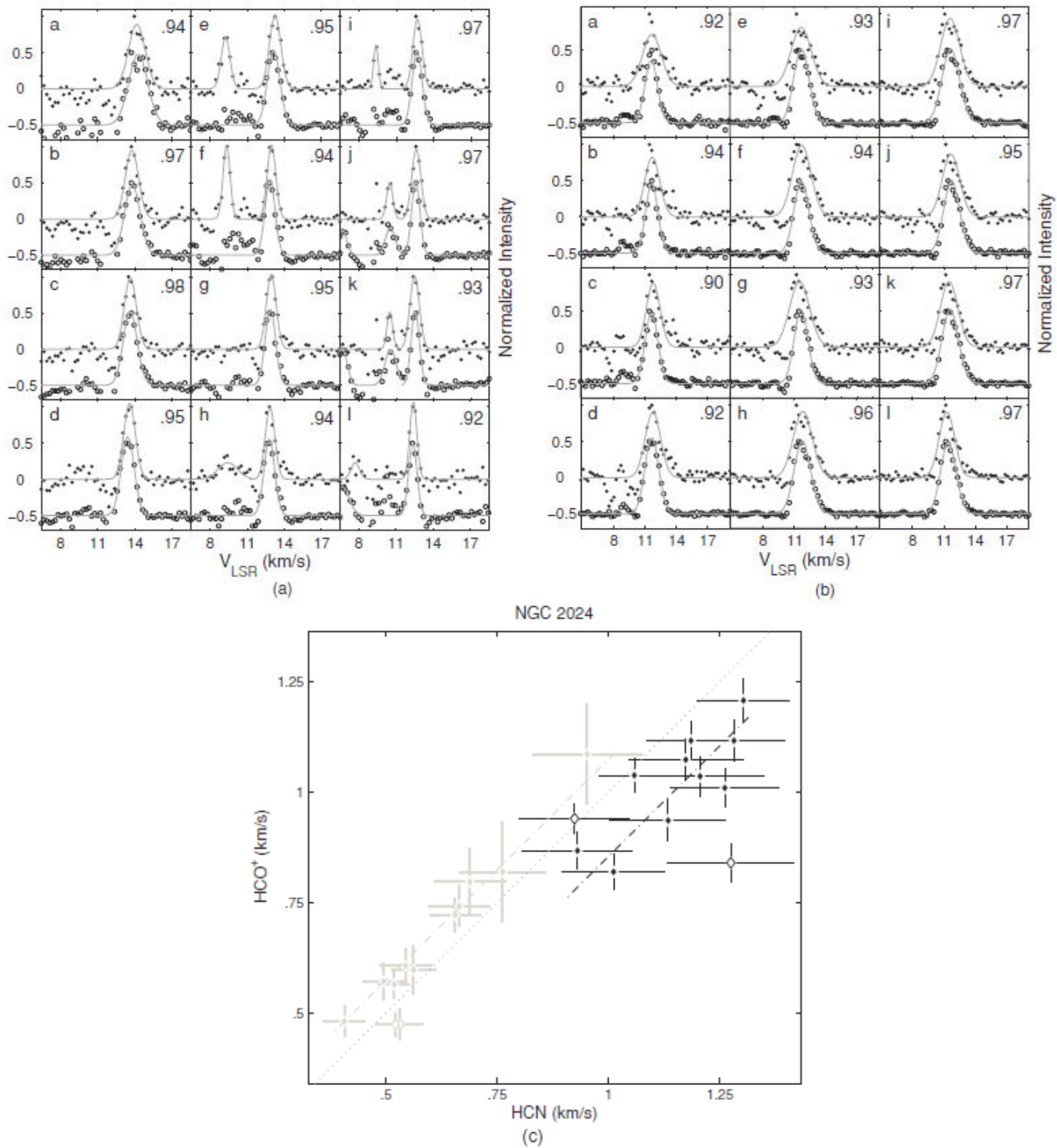

**Figure 5** (a) The spectra with normalized intensity of $HCO^+$ (o) and HCN (.) from NGC 2024 A used for VD comparison and their Gaussian fits. Note that the profile might be complicated with multiple velocity peaks, but there are always channels (11-16 km/s) that are highly correlated between the two species, and we compare VDs only for these channels. The number in the upper-right corner of each panel is the correlation coefficient of the channels used for VD comparison.
(b) Similar to (a) but for NGC 2024 B. The correlation coefficients are calculated for channels 9-15 km/s.
(c) Similar to Fig. 3b but for NGC 2024. The data are fitted without the farthest two outliers, which are denoted by the "hollow diamonds".

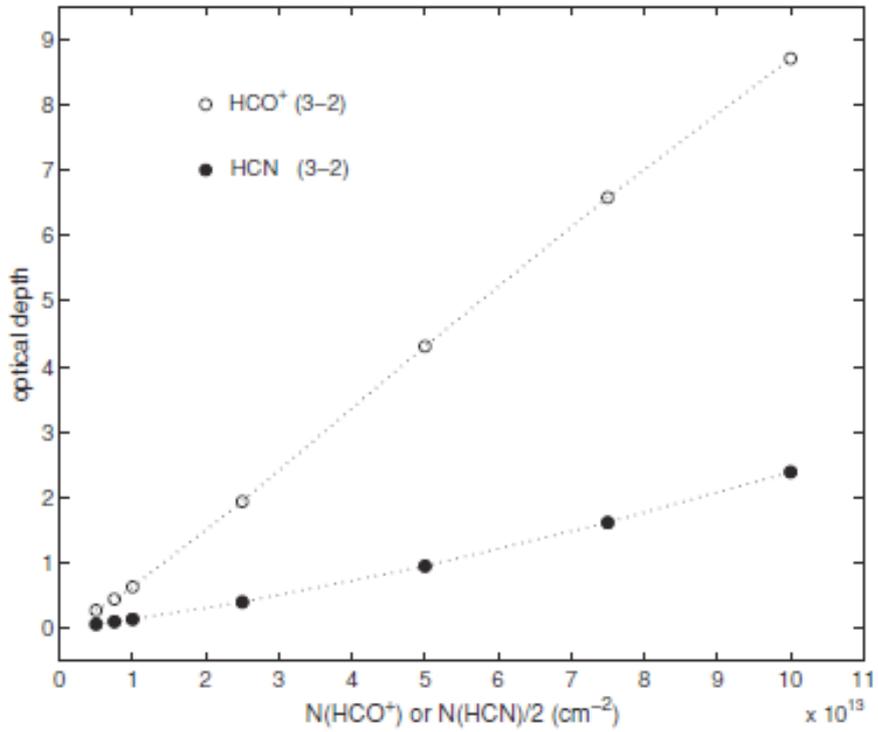

**Figure 6.** The optical depths of the HCO$^+$ and HCN (3-2) lines in NGC 2024 calculated using the RADEX program (Van der Tak et al. 2007) as a function of the column density. The input parameters are n(H$_2$)=10$^5$cm$^{-3}$, T$_K$=20K, and the typical line width of 3 km/s. The abundance ratio of HCN to HCO$^+$ is assumed to be 2, as in a typical dark cloud (Lucas & Liszt 1996).

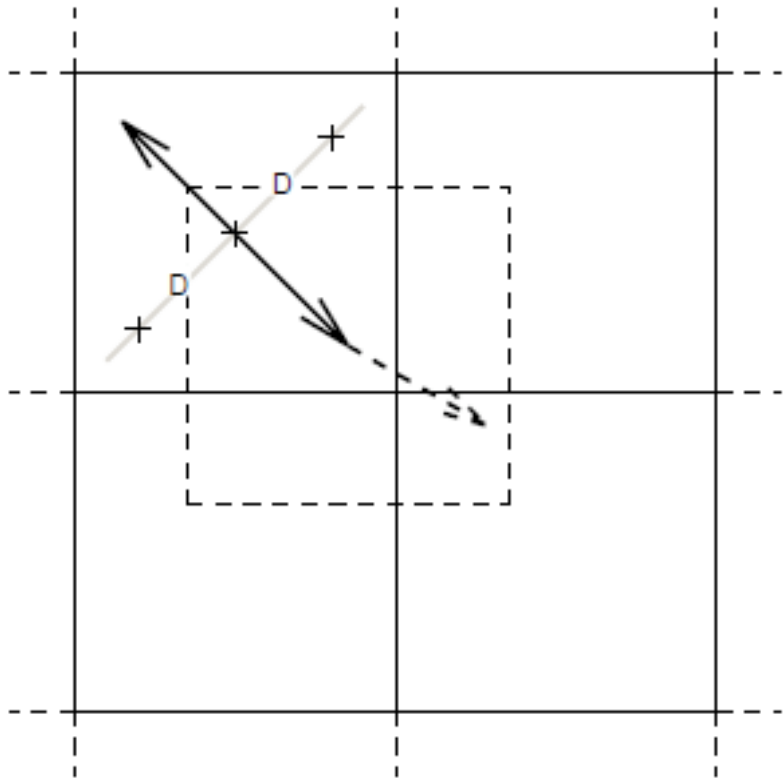

Figure 7. An illustration of how "B-line"s are plotted based on polarization data. See details in the appendix.